\documentstyle[multicol,pre,aps,epsf]{revtex}   
\begin{document}
\draft


\title{Molecular weight effects on chain pull-out fracture of reinforced 
polymeric interfaces}

\author{Mohsen Sabouri-Ghomi$^{1}$, Slava Ispolatov$^{1,2}$, and Martin
Grant$^{1}$}

\address{
$^1$Physics Department and Center for the Physics of Materials,
Rutherford Building, McGill University, 3600 rue University, Montr{\'e}al, 
Qu{\'e}bec, Canada H3A 2T8
}

\address{
$^2$Chemistry Department, Baker Laboratory, Cornell University, Ithaca,
NY 14853,USA
}

\date{\today}

\maketitle

\begin{abstract}
Using Brownian dynamics, we simulate the fracture of polymer interfaces
reinforced by diblock connector chains.  We find that for short chains
the interface fracture toughness depends linearly on the degree of
polymerization $N$ of the connector chains, while for longer chains the
dependence becomes $N^{3/2}$.  Based on the geometry of initial chain
configuration, we propose a scaling argument that accounts for both
short and long chain limits and crossover between them.  
\end{abstract}

\pacs{81.05.Qk, 62.20.Mk, 68.35.Gy}

\begin{multicols}{2}
\narrowtext
\tighten

Creating advanced materials often means mixing different homo-polymers
to produce systems with desired combined properties.  However, most
polymer blends are immiscible: They form macroscopically
phase-separated mixtures with only interfacial van der Waals forces
keeping domains of different phases together.  The fracture toughness
of such blends is limited by one of the interfaces: In ideal
defectless conditions, it is equal to the work of adhesion
$W=\gamma_{a}+\gamma_{b}-\gamma_{ab}$ between two homo-polymer phases.
Here, $\gamma_{a}$ and $\gamma_{b}$ are respectively the surface energy
of homo-polymer {\bf A} and homo-polymer {\bf B} and $\gamma_{ab}$ is
the interfacial free energy. Reinforcement of these weak polymeric
interfaces is often achieved by the addition of {\bf A}-{\bf B} diblock
copolymers which compatibilize the blend and strengthen the interface
\cite{rev0,rev1,rev2}.  The strengthening can be attributed to the
miscibility of each block with one of the homopolymers.  This causes
the block copolymer to expand and entangle with homo-polymer phases on
either side of the interface. The interfacial tension (energy) is
reduced, the interfacial width is increased, and the adhesion thereby
improved.  Fracture toughness and failure mechanisms of such reinforced
polymeric interfaces have been investigated extensively by, for
example, experiments on different incompatible systems of polymer
glasses \cite{exp1,exp2,exp3,exp4,exp5,exp6,map2} and cross-linked
networks (elastomers) \cite{exp7,exp8}.

Several theoretical models have been proposed
\cite{theo1,theo2,theo3,theo4} to explain the reinforcing effect of
connector chains in both elastomers and glassy polymers. A ``failure
mechanism map'' has been developed \cite{exp2,theo3,map1,map2} which
relates the mechanism of interface failure to the polymerization index
$N$, surface density $\sigma$ of connector chains, and the time scale
on which the deformation occurs. According to the failure map there are
three major mechanisms: (i) chain scission, which happens whenever the
stress along the connector chain becomes larger than the strength of
the covalent bond between segments of the chain, (ii) pull-out of the
connector chain as a result of disentanglement from homo-polymer phase,
and (iii) failure by craze formation, followed by chain scission or
chain pull-out, which take place when a large stress is transfered to
the bulk of the homo-polymer phases.

In this paper we direct our attention to the case where interface
failure is due to chain pull-out, and specifically focus on the effect
of the polymerization index $N$ of the connector chain.  We consider a
``mushroom regime'' \cite{Balazs,coverage1,coverage2,monte,healing,mushroom}
where connector chains are grafted with low surface
density, $\sigma \ll 1/Nl^{2}$ ($N$ is the number of monomers per
grafted chain and $l$ is the monomer size).  In this low-density
regime, the equilibrium shape of the connector chain in the homo-polymer phase 
is mushroom or plume-like.  Furthermore,
mutual entanglement between different connectors is negligible.  Hence
we need consider only the behavior of a single chain.  Experimental
data on the dependence of $G$ on $N$ is scattered, typically assuming a
form 
\begin{equation}
G\sim N^{\alpha},
\end{equation}
where estimates give variously $1\leq\alpha\leq 2$ \cite{exp2,map2}.
Both linear and quadratic dependences of $G$ on $N$ have been
predicted, utilizing a tube picture \cite{tube}.  Different
constitutive equations have been proposed which relate the local stress
in planar cohesive zone near the crack tip to various phenomena, such
as, the
penetration depth of the chains, their surface density, and the pulling
rate \cite{theo1,theo2,theo3}.  These studies have been focused mainly
on the pull-out fracture in tensile mode, when the applied force is 
normal to the interface. An alternative mode of
interface failure is shear fracture, or the resistance of the interface
against slip.  In fracture mechanics of bulk materials generally only
the tensile or opening mode is important, since cracks normally travel
in a direction that maximizes the opening mode.  However, for the
interface between two different materials, the situation can be more
complex: The crack is often constrained to follow the interface, giving
rise to the possibility of crack propagation involving a combination
of  tensile and shear modes \cite{rev1}.  Herein we study the dynamics
of chain pull-out fracture in both tensile and shear mode separately
and examine the dependence of $G$ on $N$.

We use a Brownian dynamics method, due to Picket, Jasnow, and Balazs
\cite{Balazs}, to simulate the pull-out of a single connector chain of
length $N$ from a two-dimensional homo-polymer phase.  To quantify
the interface toughness for different chain length, we calculate the
work that is required to pull out the chain with constant velocity
${\bf v_{0}}$. The homo-polymer phase is modeled by a two-dimensional
semi-infinite square lattice of obstacles.  Each obstacle represents an
entanglement or cross linked point, depending on the glassy or
elastomeric structure of the homo-polymer phase, and provides lateral
constraint on the movement of the connector chain. The connector is
represented by a freely jointed chain with $N$ links wherein there is
no self-interaction. Hence individual monomers can freely pass over
each other, provided links have a constant length.  The initial
configuration is created by putting the first monomer at the boundary
and then continuing the chain as a random walk in dimension $d = 2$,
with a reflecting boundary at the interface.  The random walk is
restricted by requiring, as we shall see below, that monomers are
repelled by the obstacles.  The pull-out dynamics is simulated by
pulling the chain by the first monomer at a constant velocity in either
tensile mode (perpendicular to the interface) or shear mode (parallel to
the interface).  This is done conveniently by moving the obstacle matrix
at velocity $ - {\bf v_{0}}$ while keeping the first monomer fixed.
That is, if 
${\bf{r_{i}}}$ is the position of the i{\it th\/} monomer,
$d {\bf{r_{1}}}/dt \equiv 0$.  

The movement of the other $i = 2,\ldots, N$ monomers
is governed by the following over-damped Langevin equation.
\begin{equation}
\nu \left(\frac{d{\bf r_{i}}}{dt}-{\bf v_{0}}\right)  =
{\tau}_{i}({\bf r_{i+1}}-{\bf r_{i}})
- {\tau}_{i-1}({\bf r_{i}}-{\bf r_{i-1}})+
{\bf F_{i}}+{\bf \eta_{i}},
\label{man}
\end{equation}
where ${\bf r_{N+1}}\equiv {\bf r_{N}}$.
Here 
${\tau}_{i}$ is the amplitude of tension in the segment connecting
monomers $i$ and $i+1$, $\eta_{i}$ is the random
Gaussian noise representing the effect of thermal fluctuations on each
monomer, and $\nu$ is the viscous friction coefficient per monomer.
The strength of the random noise is related to the monomeric friction
by the fluctuation-dissipation theorem
$\langle\eta_{i}(t)\eta_{j}(t^{'})\rangle= 2\nu
k_{B}T\delta_{ij}\delta(t-t^{'})$, where $k_{B}$ is Boltzmann's constant
and $T$ is temperature.  ${\bf F}_{i}(\Delta  {\bf r_{i}})$ is a 
short range monomer-obstacle repulsive force, 
where $\Delta  {\bf r_{i}}$
is the distance between the i{\it th} monomer and  the closest obstacle.
For $|\Delta {\bf r_{i}}| <r_c$, there is hard-core repulsion.
For $r_{c}<|\Delta {\bf r_{i}}|<2r_{c}$, there is soft-core
repulsion, where the force obeys
\begin{equation}
{\bf F_{i}} = \sigma{\Delta \bf  {\bf r_{i}}} \left( \frac{1}{|\Delta
 {\bf r_{i}}|^{2}-r_{c}^{2}}-\frac{1}{3r_{c}^{2}} \right)^{2}.
\end{equation}
The parameter $\sigma$ controls the strength of the
force \cite{Deutsch}.
For larger separations, ${|\Delta {\bf r_{i}}|>2r_{c}}$,
there is no force acting on the monomer.
In our simulations
we have used $r_{c}=l/2$, where $l$ is the link length
of the connector chain.

To solve Eq.\  (\ref{man}), the tensions ${\tau}_{i}$ in each
segment of the connector chain have to be determined. This is done by
enforcing the constant segmental length constraint:  
\begin{equation}
|{\bf r_{i+1}- r_{i}}|^{2} = l^{2},
\end{equation} 
for $ 1<i<N-1$. 
This results in a tridiagonal matrix equation for
the ${\tau}_{i}$'s \cite{Deutsch}:
\begin{eqnarray}
\frac{d( {\bf r_{i+1}}- {\bf r_{i}})^2}{2dt} & = &
{\tau}_{i+1}( {\bf r_{i+2}}- {\bf r_{i+1}})  ( {\bf
r_{i+1}}- {\bf r_{i}})-2{\tau}_{i}l^{2}+  \nonumber \\
 & + & {\tau}_{i-1}( {\bf r_{i}}- {\bf r_{i-1}})  ( {\bf r_{i
 +1}}- {\bf r_{i}})+ \nonumber  \\
 & + & ( {\bf F_{i+1}}+{\bf \eta_{i+1}}- {\bf F_{i}}-{\bf
\eta_{i}})  ( {\bf r_{i+1}}- {\bf r_{i}}).  
\end{eqnarray}
Ideally, by applying the constant segmental length constraint,  
the left-hand side of Eq.\ (4) should be zero.  However,  
after several updates of the simulation, as a result of accumulation
of roundoff errors,
the distances between some of the adjacent monomers 
differ slightly from $l$.
To improve the stability of the algorithm,
a correction, restoring the original
segment length, is introduced:
\begin{equation}
\frac{d( {\bf r_{i+1}}- {\bf r_{i}})^2}{2dt} = - 
|{\bf r_{i+1}}- {\bf r_{i}}| \cdot 
\frac{( |{\bf r_{i+1}}- {\bf r_{i}}|-l)}{dt}  .
\end{equation}
Eq.\  (\ref{man}) is solved using a fourth-order Runga-Kutta algorithm.

We consider reduced units in which $\nu$, $v_{0}$, $\sigma$, $k_{B}T$,
and $D$, the distance between obstacles, are set equal to unity.  The
distance between two adjacent monomer in connector chain $l$ is chosen
to be 0.4. The time step of the
simulation, typically $10^{-4}$ to $10^{-5}$, is adjusted so that the
average difference in segment length from $l$ in each run is less than
0.1 percent.  The results were averaged over $20-100$ independent
realizations of the initial conditions.
\vspace{-1.5cm}
\begin{figure}
\centerline{\epsfxsize=6.cm \epsfbox{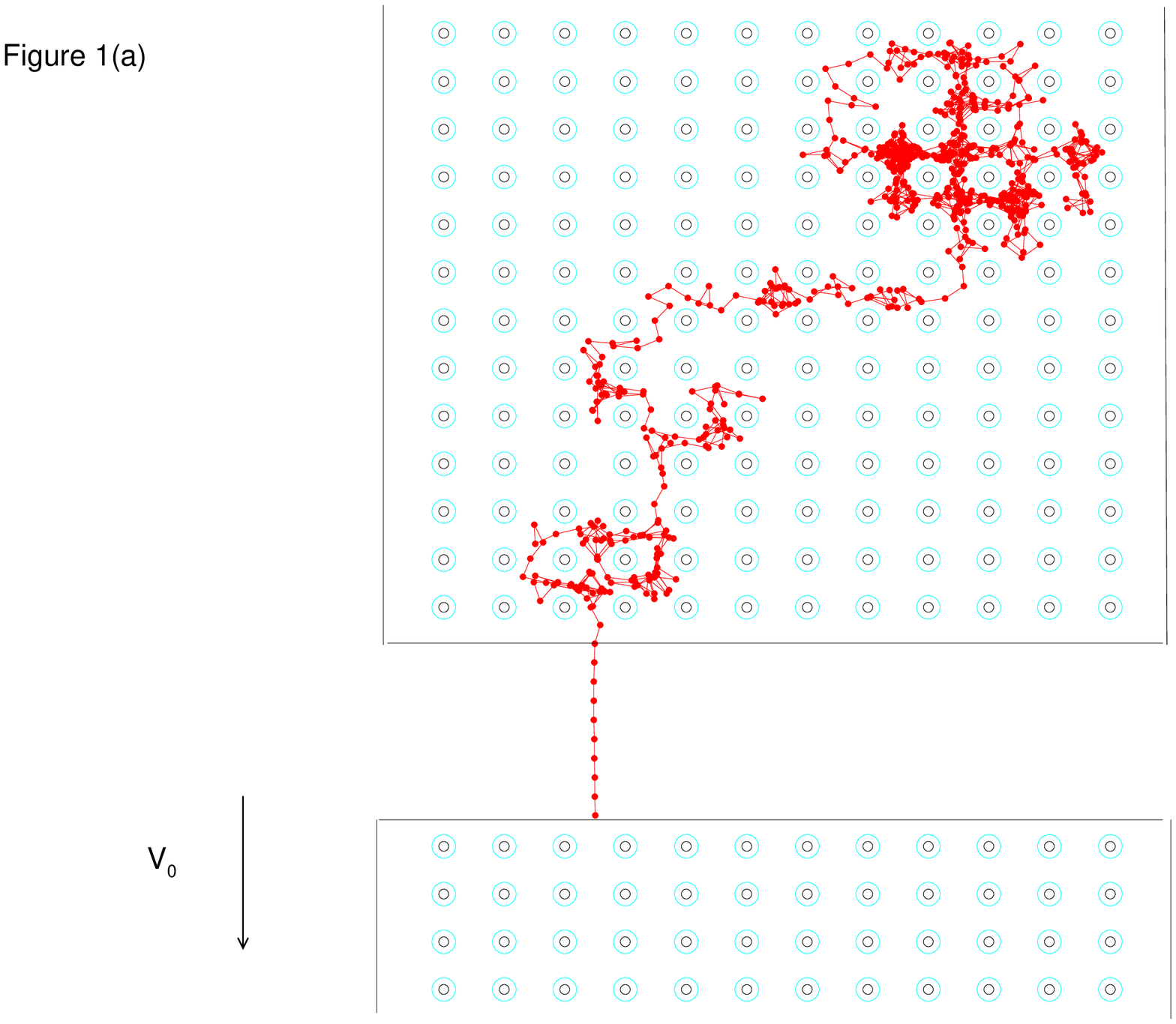}}
\centerline{\epsfxsize=6.cm \epsfbox{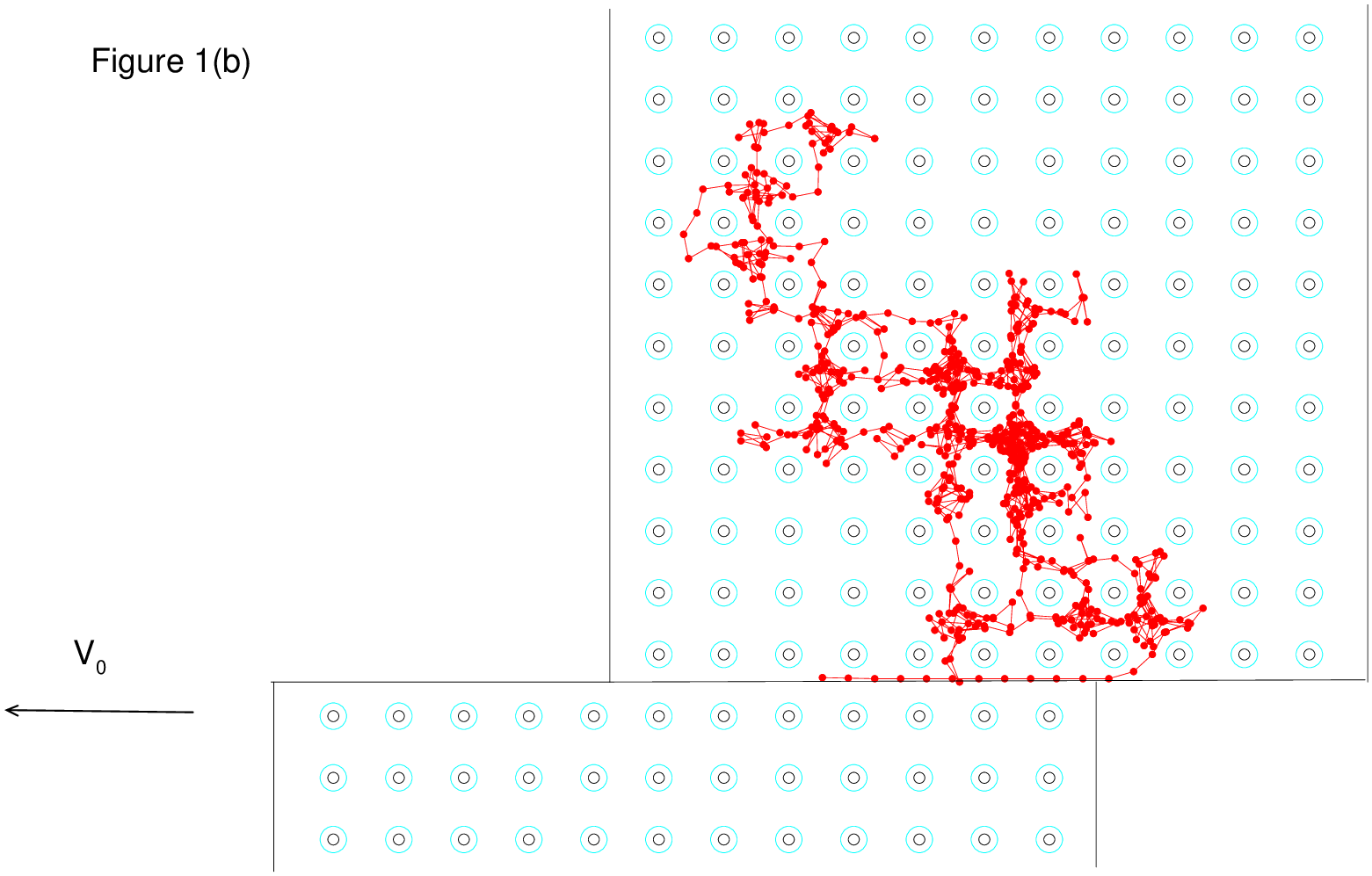}}
\caption {Snapshot of a simulation of fracture 
for a chain of $N=700$ monomers: (a) tensile and (b) shear
mode.}
\end{figure}   
Fig.\ 1 shows a snapshot of the simulation for both tensile and shear
pull out.  We quantify the
fracture toughness of the interface by determining the
work $G$  required to remove the connector chain
completely from the obstacle lattice.   The work is 
$G=\int  Pdt$, where the power $P$ is obtained via
$P= v_{0} \tau$,  
$v_0$ is the constant pulling velocity, and $\tau$ is the instantaneous
tension in the segment crossing the boundary of the obstacle lattice
in the pulling direction at each time step.
\vspace{-1.5cm}
\begin{figure}
\centerline{\epsfxsize=6cm \epsfbox{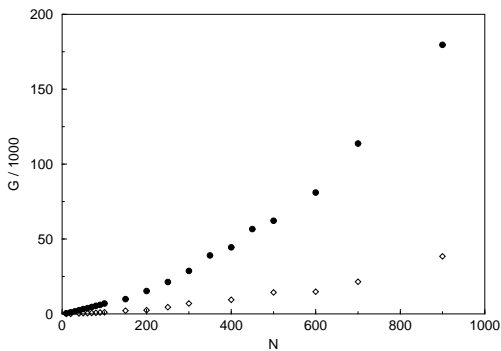}}
\caption {Ln-Ln plot of $G$ vs.\ $N$.
$(\bullet)$ represents tensile and ($\diamond$) indicates 
shear mode. Solid lines show that short-chain data ($N\leq 100$) are
consistent with $\alpha =1$, while long-chain data ($N> 100$) are
consistent with $\alpha = 3/2$.
}
\end{figure}

Fig.\ 2 shows the fracture toughness $G$ as a function of
the polymerization index $N$ of the connector chain for tensile and shear
modes.
Naturally, the fracture toughness grows as  
the degree of polymerization of the connector chain increases.
This growth is more significant for larger
$N$ due to the fact that longer connectors can penetrate well beyond the
neighborhood of the interface.  They entangle efficiently with the bulk 
polymers of the compatible phase. It also shows that the effect of
reinforcement on the interface is higher in tensile mode, by a factor of
about five for the present model \cite{noise}. 
One can notice the existence of two
scaling regimes for $G$ vs.\ $N$, corresponding to short
connectors, $N < N_{c}$, and long connectors, $N > N_{c}$ where $N_{c}$
is the crossover length.

Figure 3 shows the power-law dependence of $G$ on
$N$.  For short connectors $G$ scales linearly with $N$, $G \sim N$,
while for long connectors it scales as $G \sim N^{3/2}$
\vspace{-1.5cm}.
\begin{figure}
\centerline{\epsfxsize=6cm \epsfbox{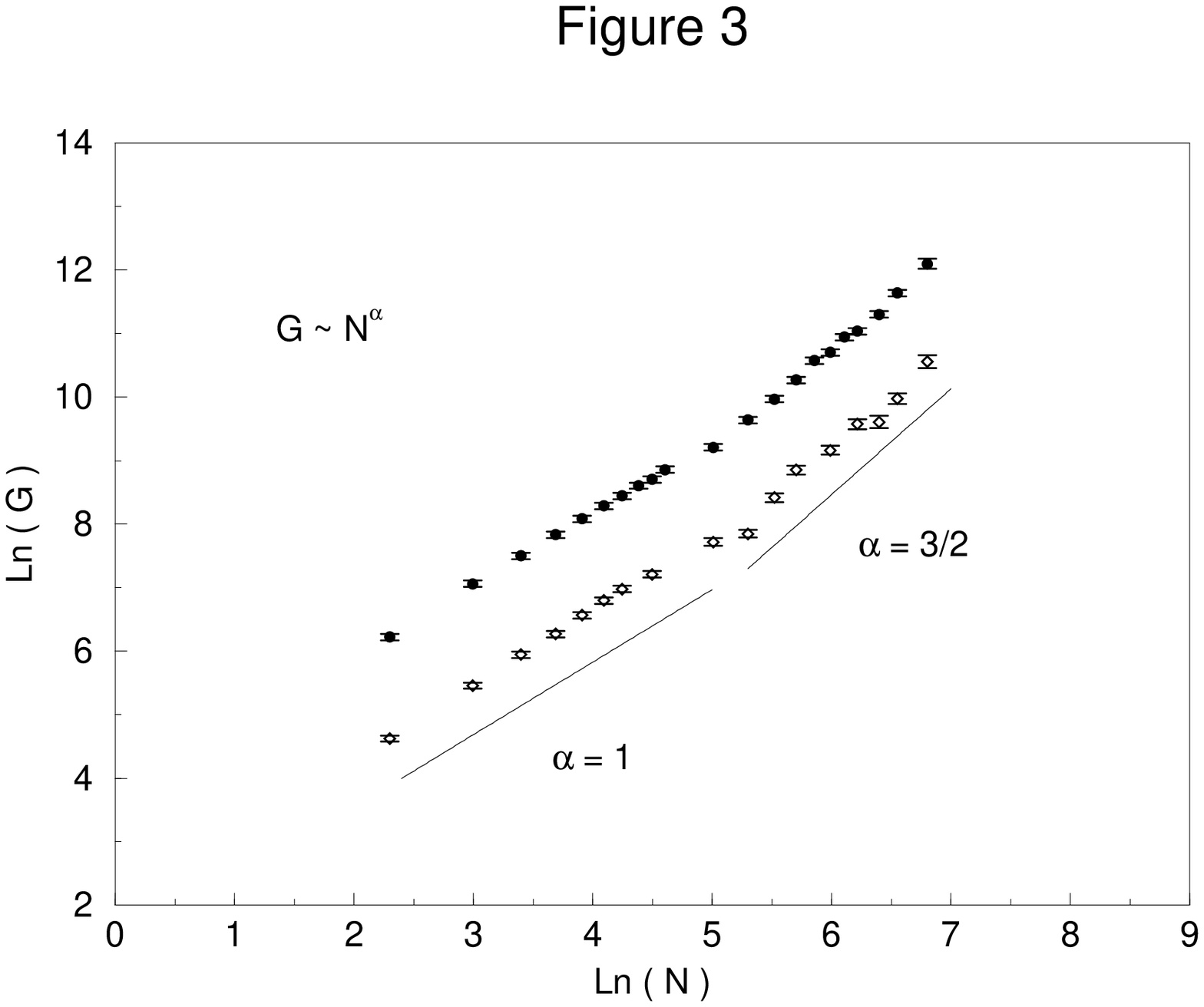}}
\caption {Log-log plot of $G$ vs.\ $N$ for long connectors $N=150$ to 
$900$.
($\bullet$) and ($\diamond$) represent tensile and shear modes respectively.    
Data is consistent with $G \sim N^{3/2}$ (solid line).
}
\end{figure}

These regimes can be understood as follows.
For small $N$, the chain is
entangled with one or two layers of obstacles, even if
the free radius of gyration $R_{g}\approx 
l\sqrt{N}$ is of order
or smaller than the obstacle lattice spacing $D$.
Since the 
the initial configuration is repelled by 
the obstacles, 
the available phase space for a chain is quite restricted.
For the first few chain segments, the reflecting boundary condition and the 
the area taken up by the obstacles,
results in
the chains going straight into the lattice. After passing the first
row or two of obstacles, the chain, if it is long enough,  
usually bends, with
tight loops around obstacle cores being highly unprobable.
When the chain is pulled out, the dominant deterministic
part of the tension equals the total viscous drag force.
This force is roughly proportional to
the number of monomers $N_{move}$
simultaneously in motion.  Since pulling tension cannot propagate 
through a loop unless it
is tightened around an obstacle core, for short chains 
$N_{move}$ is of order of
the number of monomers stretched in one obstacle lattice spacing $D$,
that is $D/l$. 
Hence for small $N$, the total work to pull the chain out
can be estimated as 
\begin{equation}
G \approx {v_{0}}^2 \nu {D\over l} \int_{0}^{Nl/{v_{0}}}dt 
=  { v_{0}}\nu N D, 
\end{equation}
which is consistent with the linear scaling regime observed in simulations
for short connectors.

For large $N$,  the penetration depth increases.  Indeed,
for sufficiently large $N$, the chain center of mass is
located at a distance of order of the radius of
gyration $R_{g}$ from interface. Then, the number
of monomers that are simultaneously in motion becomes
proportional to this distance, $N_{move} \approx R_{g}\sim N^{1/2}$,
which we have observed directly.
Consequently, the pull-out fracture energy becomes
\begin{equation}
G = {v_{0}}\int_{0}^{Nl/{v_{0}}} F_{drag} dt \approx 
v_{0} \nu N^{3/2} l.
\end{equation}
This is roughly analogous to what happens when one pulls on a wet
garden hose of length $N$, left on the ground by a
gardener who has performed a random walk among a grove of trees.
Upon pulling, only the currently stretched segment starts to move,
if friction in the self-intersections of the hose is
negligible. If the hose configuration was created as a result of 
random walk, the length of such a segment scales as 
a typical size of such a walk, $\sqrt{N}$, where
$N$ is the total length of the hose.  The hose configuration may
include a few loops around the tree trunks, which, upon tightening, also
start to move. The total length of these tightened loops has no
important contribution:  That length scales as a
winding angle of a random walk of the length $N$, {\it i.e.\/}, as
$\ln N$ (see, for example, Ref.\ \cite {drossel}). As a result, the drag
force is proportional to the length of the hose that is constantly in
motion, $\sqrt{N}$.

Although our numerical
work is in two dimensions, these physical arguments also follow in three
dimensions, and are applicable to experimental systems
\cite{Nsquared}.
The small $N$ regime crosses over to a large $N$ regime when
the radius of gyration $R_g$ becomes of order of the 
obstacle spacing $D$.  
That gives $N_{cross}\sim [D/l]^2$,
and we expect 
\begin{equation}
G = N f(N/N_{cross}),
\end{equation}
where the crossover scaling function obeys
$f(x\rightarrow 0) = {\rm const}$, and 
$f(x\rightarrow \infty) = x^{1/2}$.
  
As mentioned earlier, different forms of the power-low dependence of $G$ on
$N$ have been proposed in previous theoretical studies: Both $G
\sim N$ and $G\sim N^{2}$ have been predicted.  The 
results from experiments are 
somewhat ambiguous. The linear dependence of $G$ on $N$
\cite{theo1} is predicted for very slow crack propagation velocities, 
and therefore
corresponds to $G \rightarrow G_{0}$,  
where $G_{0}$ is the fracture toughness threshold or minimal
energy required to break the interface when $ {v_{0}} \rightarrow
0$. On the other hand, $G\sim N^{2}$ is predicted for pull-out fractures
where the crack propagation speed is  high or at least larger than a critical 
velocity.  In other words, these two
predicted regimes differ by the pull-out velocity. 
The crossover between the two scaling regimes $G\sim N$ and $G\sim N^{3/2}$  
obtained in our simulations takes place in
the latter high velocity regime: The pulling
velocity used in our simulations $ v_0  = k_BT/(\nu D)$
corresponds to relatively fast pulling rate \cite{Ajdari}, which   
in turn results in high crack propagation speeds. 
Our work is in the same regime, and indeed 
extends the numerical work and theoretical arguments of 
Picket, Jasnow, and Balazs \cite{Balazs}, wherein this model was
introduced.  For large $N$, Picket {\it et al.\/}\ argued 
$G \sim N^{2}$ in contrast to our result of $G \sim
N^{3/2}$.    This was based on an  analogy of the polymer to 
the motion of a rope in ``block and tackle" pulleys.  In that case, the
average drag force to pull the connector chain out of the matrix
is $f_{drag} \sim \nu
v_0 N$.  Hence, it follows that the number of monomers instantaneously
in motion are $N_{move}\sim N$.  In fact, the correct analogy is not to
pulleys, but to 
the motion of a garden hose lying on grass, as described above.  
Then $f_{drag} \sim N^{1/2}$, giving rise to the $3/2$ exponent.
Our numerical results, which are more extensive than those of the
earlier work, support this picture.

To conclude, we studied the chain pull-out fracture of a reinforced
polymeric interface in tensile and shear modes using molecular dynamics
algorithm. Our results confirm the nonlinear dependence of the fracture
toughness of the interface on the length of the connector chain
observed in experiments.  We found that, depending on the length of
connector chain, the fracture toughness of the interface shows
different scaling dependences: For short chains, $G$ scales linearly
with $N$, while for long connectors, we observed a crossover to a new
scaling regime $G \sim N^{3/2}$.  Our results can be tested
experimentally on elastomeric networks for long connector chains in the
low coverage mushroom regime \cite{EXPERIMENT}.

This work was supported by the Natural Sciences and Engineering
Research Council of Canada, and {\it le  Fonds pour la Formation de
Chercheurs et l'Aide \`a la Recherche de Qu\'ebec\/}.

\end{multicols}

\begin{references}
\bibitem{mushroom}
The degree of the coverage of the interface by the grafting chains
plays an important role. Neutron diffraction measurements
\cite{coverage1}, and adsorption kinetic studies \cite{coverage2}
together with Monte Carlo simulations \cite{monte} of configurational
properties of grafted diblock copolymers reveal a crossover from
essentially unperturbed ``mushroom'' conformation at low surface
coverage to strongly stretched ``brush'' conformation at high surface
coverage.  In our simulations we consider the low coverage regime where
the connector chain takes a non-biased random mushroom-like
configuration. It is consistent with the observed equilibrium
configuration of a single chain grafted to a reflective wall in contact
with an elastomeric network \cite{healing,coverage1}.
\bibitem{noise} The difference in fracture toughness in
shear and tensile modes is due to the effect of thermal noise on the
tension of the crossing bond of the connector chain at the interface. In
the shear mode, the crossing bond is always anchored by one of the obstacles in
the boundary of the obstacle lattice.  Therefore most of the thermal noise
on the crossing bond is absorbed by the soft potential shell around the
anchoring obstacle. This results in a reduction of the tension in the crossing
bond and, hence in the fracture toughness.
\bibitem{Nsquared}
If the initial configuration is created including a small bias in the
random walk, then, for extremely large $N$ 
the chain length becomes proportional to $N$ itself, giving  
$G \sim N^{2}$.
\bibitem{EXPERIMENT}
Experimental measurements of fracture toughness
of reinforced glassy and elastomeric polymer interfaces by diblock
copolymers have been carried for connector chains with very high
polymerization index ($N$ up to 870 for glassy \cite{exp2} and $N$ up
to 2205 for elastomeric \cite{exp8} interfaces).  Nevertheless, 
there are only two sets
of experimental data available \cite{exp2,map2} on the power law
dependence of fracture toughness $G$ on polymerization index $N$.  These
are obtained from experiments on glassy interfaces 
for connector chains of $N\leq 270$ with brushy initial
configurations.
\end{references}
\end{document}